\documentclass{ws-procs9x6-notrim}

\def\bfnabla{\mbox{\boldmath $\nabla$}}

\def\bfsigma{\mbox{\boldmath $\sigma$}}

\def\lQ{\Lambda_{\rm QCD}}
\def\em{{\rm em}}

\newcommand{\nn}{\nonumber}
\newcommand{\be}{\begin{equation}}
\newcommand{\ee}{\end{equation}}
\newcommand{\bea}{\begin{eqnarray}}
\newcommand{\eea}{\end{eqnarray}}

\def\als{\alpha_{\rm s}}
\def\siml{{\ \lower-1.2pt\vbox{\hbox{\rlap{$<$}\lower6pt\vbox{\hbox{$\sim$}}}}\ }} 
\def\simg{{\ \lower-1.2pt\vbox{\hbox{\rlap{$>$}\lower6pt\vbox{\hbox{$\sim$}}}}\ }}

\begin{document}

\title{RADIATIVE TRANSITIONS AND THE QUARKONIUM MAGNETIC MOMENT}

\author{ANTONIO VAIRO}

\address{Dipartimento di Fisica dell'Universit\`a di Milano and INFN, \\ 
via Celoria 16, 20133 Milano, Italy \\
E-mail: antonio.vairo@mi.infn.it}

\begin{abstract}
I discuss heavy quarkonium radiative transitions and the related issue of 
the quarkonium magnetic moment inside effective field theories. 
Differences in set up and conclusions with respect to typical phenomenological 
approaches are outlined. 
\end{abstract}


\bodymatter

\vspace{1cm}

Heavy quarkonium radiative transitions have been studied for long time 
in the framework of phenomenological models\cite{Feinberg:1975hk,Sucher:1978wq,Eichten:1978tg,Kang:1978yw,Sebastian:1979gq,Karl:1980wm,Grotch:1982bi,Moxhay:1983vu,McClary:1983xw,Grotch:1984gf,Fayyazuddin:1993eb,Lahde:2002wj,Ebert:2002pp,Barnes:2005pb}
(for a recent review see\cite{Brambilla:2004wf}).
In a typical set up\cite{Grotch:1982bi}, the starting point is a relativistic Hamiltonian whose 
dynamical fields are heavy quarks and photons; gluons are integrated 
out and replaced by a scalar and a vector interaction. The non-relativistic
nature of heavy quarkonia is exploited by performing a
Foldy--Wouthuysen transformation. Anomalous magnetic
moments are added as if the system was made of two  
particles independently interacting with the external electromagnetic field. 

Such an approach, despite phenomenological success, has an obvious 
limitation: its connection with QCD is unclear. More precisely, 
while it relies on a systematic (non-relativistic) expansion to describe 
the electromagnetic interaction of the heavy quarks, it does not to describe 
their strong interaction. As a result, the nature of the binding potential 
between the two heavy quarks and of the couplings of the heavy quarks 
with the electromagnetic field remains elusive. In particular, one may wonder if the 
heavy quarkonium  anomalous magnetic moment gets contributions from 
the interaction of the heavy quarks in the bound state 
and if, in the non-relativistic expansion, the many couplings between the heavy 
quarks and the electromagnetic field, allowed by the symmetries of QCD, really organize  
as if they were generated by only two relativistic form factors (i.e. the
scalar and vector interactions). Finally, the lack of a systematic expansion reflects 
in the fact that model-dependent determinations are affected by unknown uncertainties.

In the modern framework of non-relativistic effective field theories (EFTs)\cite{Caswell:1985ui,Bodwin:1994jh,Pineda:1997bj,Brambilla:1999xf,Brambilla:2004jw}, 
one takes advantage of the small heavy quark velocity $v$
to organize calculations of heavy quarkonium observables so 
that theoretical uncertainties can be evaluated and systematically 
reduced. Applied to radiative transitions, the EFT approach provides eventually  
definite answers to the above questions\cite{Brambilla:2005zw}. 
This is done through the following step.

First, one has to identify the relevant scales in the system.
These are the heavy quark mass $m$, the typical momentum transfer  
in the bound state, which is of order $mv$, and the typical binding energy,
which is of order $mv^2$. The momentum of the emitted photon, $k_\gamma$, 
is of order $mv^2$ for transitions between different radial levels 
(like magnetic-dipole hindered transitions) and usually of order $mv^4$ for transitions 
between levels with the same principal quantum number (like magnetic-dipole allowed transitions). 
Since the typical distance between the two heavy quarks is $r \sim 1/(mv)$,
then $k_\gamma \, r \ll 1$ and the external electromagnetic field 
can be multipole expanded (at least for transitions between not too far away levels).
We neglect virtual photons, whose contributions are  
suppressed by $\alpha$. In QCD, it is also crucial to establish the size of 
these scales with respect to the typical hadronic scale $\lQ$. By definition
of heavy quark: $m\gg \lQ$; however, $mv \gg \lQ$ holds possibly only for the lowest
quarkonium resonances, while $mv \sim \lQ$ holds for the others. 
In the first case, called weakly coupled, the scale 
$mv$ may be treated in perturbation theory (in this case $v \sim \als$), in the last one, called
strongly coupled, at the scale $mv$ one cannot rely on an expansion in $\als$ 
and non-perturbative techniques have to be used.

In order to construct an EFT suitable to describe heavy quarkonium 
transitions, we do not need to resolve scales larger than or of the same order 
as the soft scale $mv$. These scales may be integrated out from the EFT. The relevant degrees of
freedom differ in the weakly and in the strongly coupled case. 
If $mv \gg \lQ$ these are singlet, $\rm S$, and octet, $\rm O$, quarkonium fields, gluons 
of energy and momentum $mv^2$ or $\lQ$ (sometimes also called ultrasoft) and 
photons. If $mv \sim \lQ$, at scales lower than $mv$ colour confinement 
sets in and the relevant degrees of freedom are singlet quarkonium fields 
(all gluonic excitations between heavy quarks develop a mass gap of order 
$\lQ$ with respect to the lowest state and are integrated out) and photons. 
Fields scale in accordance to the lowest, still dynamical, energy scales, 
defining in this way the power counting of the EFT. 

The EFT Lagrangian has the following form:
\bea
{\cal L}
&=&  
-\frac{1}{4}   F_{\mu\nu}^{\rm em} F^{\mu\nu\,{\rm em}} 
+ \int d^3r\, {\rm Tr} \left\{ {\rm S}^\dagger \left( i\partial_0 - \frac{{\bf p}^2}{ m} 
- {V_s} \right){\rm S}\right\} 
\nn \\
& &
\left[
-\frac{1}{4}    F_{\mu\nu}^a F^{\mu\nu\,a} 
+  \int d^3r\, {\rm Tr} \left\{ {\rm O^\dagger}\left( i{D_0} - \frac{{\bf p}^2}{m} - {V_o} 
\right){\rm O}\right\} \right.
\nn \\
& &
+ \int d^3r\, V_A \; {\rm Tr} \left\{  {\rm O^\dagger}{\bf r}\cdot g {\bf E}\,{\rm S}
+ {\rm S^\dagger} {\bf r} \cdot g {\bf E} \,{\rm O}\right\} 
\nn\\
& &
\left.
+ \int d^3r\, V_B \; \frac{1}{2} {\rm Tr} \left\{  {\rm O^\dagger}
{\bf r}\cdot g {\bf E}\,{\rm O}
+ {\rm O^\dagger} {\rm O }{\bf r}\cdot g{\bf E} \right\}
+ \cdots \right]_{\hbox{weak coupling only}}
\nn\\
& &
+ {\cal L}_{\gamma}, 
\eea
where ${\cal L}_\gamma$ is the part of the Lagrangian that describes the
interaction with the electromagnetic field. In the case of magnetic-dipole 
transitions, the relevant terms in ${\cal L}_\gamma$ up to relative order 
$v^2$ are:
\bea
{\cal L}_{\gamma}  &=& \int d^3r\, {\rm Tr} \, \Bigg\{ 
V_A^\em \; {\rm S}^\dagger {\bf r}\cdot e {\bf E}^{\em} {\rm S} 
+ \frac{1}{2 m}
\; V_1
\; \left\{{\rm S}^\dagger , \bfsigma \cdot e {\bf B}^{\em}\right\} {\rm S} 
\nn\\
&& 
+ \left[\frac{
1}{2 m}
\; V_1
\; \left\{{\rm O}^\dagger , \bfsigma \cdot e {\bf B}^{\em} \right\} {\rm
  O}
\right]_{\hbox{weak coupling only}} 
\nn\\
&& 
+ \frac{1}{4 m^2}
\; \frac{V_2}{r}
\; \left\{{\rm S}^\dagger , \bfsigma\cdot\left[ \hat{\bf r} \times  \left(
  \hat{\bf r}\times e {\bf B}^{\em} \right) \right] \right\} {\rm S} 
+ \frac{1}{4 m^2}
\; \frac{V_3}{r} 
\; \left\{{\rm S}^\dagger , \bfsigma \cdot e {\bf B}^{\em}\right\} {\rm S} 
\nn\\
&& 
+ \frac{
1}{4 m^3} 
\; V_4 \; \left\{ {\rm S}^\dagger , \bfsigma \cdot e {\bf
  B}^{\em} \right\} \bfnabla_r^2 {\rm S} 
+ \cdots \Bigg\},
\label{gammapNRQCD:Lag}
\eea
where $e$ is the electric charge of the heavy quark.
The Wilson coefficients $V_s$, $V_o$, $V_A$, $V_B$, $V_A^{\em}$, $V_i$ are in general functions of $r$
and have to be determined by matching amplitudes calculated in
the EFT with amplitudes calculated in QCD. The matching may be performed order 
by order in $\als$ in the weak-coupling case, but has to be performed
non-perturbatively in the strong-coupling one.
In both situations, the matching can be done in two steps: 
the first one consists of integrating out hard 
modes of energy of the order of the heavy quark mass, the second one of
integrating out soft modes of energy of the order of $mv$. 
The Wilson coefficients of the EFT will then have the factorized form: (hard)$\times$(soft). 
For determinations of $V_s$, $V_o$, which are the singlet and octet
potentials, and $V_A$, $V_B$ we refer to\cite{Brambilla:2004jw} and 
references therein. The coefficients $V_A^{\em}$ and $V_i$ 
have been studied in\cite{Brambilla:2005zw}. In the following, we summarize 
the main conclusions.

The coefficient $V_1$ may be interpreted as the heavy quarkonium magnetic moment.
Its hard part is simply the sum of the heavy quark magnetic moments: 
$1 + 2\als/(3\pi) + \dots$. Moreover, it can be shown that, in the SU(3)$_{\rm flavour}$ limit,  
to all orders in the strong-coupling constant, $V_1$ does not get soft contributions.
The argument is somewhat similar to the factorization of the QCD corrections in $b
\to u \, e^- \bar{\nu}_e$, which fixes the coefficient in the effective
Lagrangian ${\cal L}_{\rm eff}=-4G_F/\sqrt{2}\; V_{ub} \; \bar{e}_L\gamma_\mu\nu_L\, \bar{u}_L\gamma^\mu b_L$
to be 1 to all orders in $\als$ (see, for example,\cite{Neubert:2005mu}).
This leads to the conclusion that the heavy quarkonium
magnetic moment, if defined as the Wilson coefficient $V_1$, 
is just the sum of the heavy quark magnetic moments. Therefore, its anomalous 
part is small and positive and  does not get any large low-energy and, 
in particular, non-perturbative contribution. 
This is consistent with a recent lattice determination\cite{Dudek:2006ej}.
Note that a large negative heavy quarkonium anomalous magnetic moment
has been often advocated in potential models to accommodate the
results with the data.

Reparametrization and Poincar\'e invariance\cite{Manohar:1997qy,Brambilla:2003nt} protect the coefficients $V_2$ and
$V_3$. To all orders in perturbation theory and non-perturbatively it holds
that $V_2 = r^2 V^{(0)\,\prime}_s/2$ ($V^{(0)}_s$ is the singlet static potential)
and $V_3=0$. Since an effective scalar interaction 
would contribute to $V_3$, we conclude that such an interaction is not 
dynamically generated in QCD for magnetic-dipole transitions.
Again, a scalar interaction is often employed in potential models.

The coefficient $V_A^{\em}$ is 1 up to possible corrections of order
$\als^2$. In general, the coefficients of the operators of order $1/m^3$ are 
not protected by any symmetry. $V_4$ is 1 plus ${\cal O}(\als)$ 
corrections. In the weak-coupling case, this is the only $1/m^3$ operator
needed to describe magnetic-dipole transitions at relative order $v^2$. 
In the strong coupling regime, more terms are, in principle, necessary.
Since $\als(1/r)\sim 1$ is no longer a suppression factor, 
more amplitudes shall contribute to the matching.
These amplitudes will be encoded in the matching coefficients of
the EFT in the form of static Wilson loop amplitudes with  
field strength insertions of the same kind as those that appear in the QCD potential 
at order $1/m^2$ \cite{Pineda:2000sz}. Also, they may induce new operators in
the EFT. A non-perturbative derivation of the EFT Lagrangian coupled 
to the electromagnetic field at order $1/m^3$ has not been worked out yet; 
note that to be of phenomenological impact such a calculation 
needs to be supplemented by lattice calculations of the relevant Wilson loop amplitudes.
Since most of the potential model calculations rely on Lagrangians 
of the type of Eq.  (\ref{gammapNRQCD:Lag}), we emphasize that, once cleaned
up of the scalar interaction, they are suitable to describe 
the relativistic corrections for radiative transitions between weakly-coupled 
quarkonia ($mv^2\simg \lQ$) only. This is a rather severe constraint since most 
of the radiative transitions (e.g. magnetic dipole hindered, electric dipole,
...) involve higher excited quarkonium states, which may be difficult to
accommodate as Coulombic bound states. 

At present, allowed magnetic-dipole transitions 
between the quarkonium lowest states are those most reliably described in QCD.
Moreover, it has been shown in\cite{Brambilla:2005zw} that for these
processes, non-perturbative contributions mediated by quarkonium octet states and ultrasoft 
gluons cancel at relative order $v^2$. Therefore, at relative order $v^2$, 
transitions like $J/\psi \to \eta_c \, \gamma$ and $\Upsilon(1S) \to \eta_b\, \gamma$ 
are completely accessible in perturbation theory. An explicit calculation
gives:
\bea
&& \Gamma(J/\psi \to \eta_c\,\gamma) = (1.5 \pm 1.0)~\hbox{keV},
\label{M1cc}\\
&& \Gamma(\Upsilon(1S) \to \eta_b\, \gamma) =
\left(\frac{k_\gamma}{39~\hbox{MeV}} \right)^3 \, 
(2.50 \pm 0.25)~\hbox{eV}.
\label{M1bb}
\eea
The error in Eq. (\ref{M1cc}) accounts for the large uncertainties coming from
higher-order corrections; we recall that corrections of order
$k_\gamma^3\,v^2/m^2$ affect the leading-order result
by about 50\%. Uncertainties may be reduced by higher-order calculations. 
The value given in Eq. (\ref{M1cc}) is  consistent with the 
present experimental one\cite{Yao:2006px}. 
This means that assuming the ground-state charmonium 
to be a weakly-coupled bound state leads to relativistic corrections to the 
transition width of the right sign and size. 
The decay width $\Gamma(\Upsilon(1S) \to \eta_b\, \gamma)$ depends on the
$\eta_b$ mass, which is unknown. The quoted value corresponds to a $\eta_b$
mass of about 9421 MeV.

We conclude with few possible developments of the work discussed here.
For magnetic-dipole transitions between higher resonances, the completion of
the non-perturbative matching of the relevant operators in the EFT at order
$1/m^3$ will be needed, possibly integrated by lattice calculations 
of the relevant Wilson-loop amplitudes. 
For magnetic-dipole hindered transitions of the type $\Upsilon(3S)\to \eta_b\gamma$,
since the momentum of the emitted photon is comparable with the typical
momentum transfer in the bound state, one cannot rely on the multipole expansion of the
external electromagnetic field. In this case, one may, in principle,  exploit
the fact that the typical  momentum transfer inside the $\eta_b$ is much
larger than that one inside the $\Upsilon(3S)$. A suitable treatment has not 
been developed yet. For electric-dipole transitions, much of the study 
still remains to be done.  In the weak-coupling regime, octet contributions
may not vanish; they can be worked out as in the case of 
the magnetic-dipole transitions. However, most of the electric-dipole transitions 
may need to be treated in a strong-coupling framework.

\section*{Acknowledgements}
I thank Nora Brambilla and Yu Jia for an enjoyful collaboration 
on the work presented here. I acknowledge the financial support obtained inside the Italian
MIUR program  ``incentivazione alla mobilit\`a di studiosi stranieri e
italiani residenti all'estero''.

\end{document}